# Highly efficient anomalous refraction of airborne sound through ultrathin metasurfaces

Kun Tang, Chunyin Qiu*, Manzhu Ke, Jiuyang Lu, Yangtao Ye, and Zhengyou Liu*

Similar to their optic counterparts, acoustic components are anticipated to flexibly tailor the propagation of sound. However, the practical applications, e.g. for audible sound with large wavelengths, are frequently hampered by the issue of device thickness. Here we present an effective design of metasurface structures that can deflect the transmitted airborne sound in an anomalous way. This flat lens, made of spatially varied coiling-slit subunits, has a thickness of deep subwavelength. By elaborately optimizing its microstructures, the proposed lens exhibits high performance in steering sound wavefronts. The experimental results are in excellent agreement with the theoretical predictions. This study may open new avenues for numerous daily life applications, such as controlling indoor sound effects by decorating rooms with light metasurface walls.

Key Laboratory of Artificial Micro- and Nano-structures of Ministry of Education and School of Physics and Technology, Wuhan University, Wuhan 430072, China

*Correspondence and requests for materials should be addressed to C.Q. (email: cyqiu@whu.edu.cn) or Z.L. (email: zyliu@whu.edu.cn).

It is well-known that optic components (OCs) can be flexibly designed to control light propagation by gradually tailoring phase fronts, which leads to a great number of practical applications. Comparatively, acoustic components (ACs) have received much less attention, although they are expected to undertake similar functionalities in acoustics. The dilemma is mainly originated from two critical factors that limit the performance of the conventional ACs, especially for airborne sound in the audible regime that is closely related to our daily life.

The first major limitation stems from the acoustic opaqueness of natural solids for airborne sound, due to the extreme impedance contrast with respect to air. The opaqueness strongly suppresses the transmission efficiency of ACs. This drawback is considerably relaxed by the recent progress on the air-based artificial structures. In 2002 Cervera *et al.* have reported[1] that sonic crystals can be used to design various acoustically transparent refractive devices at the frequency of the first band. For the higher frequency bands, sonic crystals have also been proposed to fabricate planar lens based on the fascinating negative refraction effect[2-4]. Intuitively, the excellent transparency in sonic crystals stems from the substantially improved impedance-matching due to the existence of air channels for direct sound propagation. Unwanted reflections from such devices can be further reduced by attaching carefully designed anti-reflection layers[5]. Transparency can also be realized in acoustic metamaterials[6-20], which often closely connects with resonant acoustic responses[6-10] or strong anisotropies[11-13,15-20] of subwavelength units. The unnatural sound responses endow the metamaterials with unprecedented capabilities in tailoring sound, such as subwavelength imaging[11-16] and cloaking[18-20].

The second barrier to the high performance of ACs is their notable thicknesses. Similar to the conventional OCs, the thicknesses of ACs are often much larger than the wavelength of operation. This severely restricts the miniaturization of ACs, especially for low frequency (e.g. audio range). This issue cannot be solved by the artificial materials based on bulk effects as well. Recently, the two-dimensional (2D) equivalent of the metamaterial, i.e. the so-called metasurface structure (MS), has attracted a tremendous interest in the optics community[21-27]. Yu and coworkers[21] have demonstrated an unusual manipulation of light wavefronts through an ultrathin MS, where the reflection or refraction waves are redirected and follow the so-called generalized Snell's law (GSL). The anomalous wavefront redirection is accomplished by designing a constant gradient of the phase accumulation over a flat layer decorated with spatially varying plasmonic units. In terms of physics, the momentum mismatch between the incident wave and the deflected wave is compensated by the MS-induced transversal momentum. Based on a similar principle, the flat MS can even reshape light wavefronts in nearly arbitrary ways providing that appropriate 2D spatial phase profiles are molded[21-25]. Comparing to the conventional OCs, the ultrathin property enable the MSs to be more compatible with on-chip nano-photonic devices, which is of significant importance for future applications. Based on the surface equivalence principle[26] or the optical nanocircuit concept[27], alternative design routes have been further proposed to improve the coupling efficiency to the desired transmitted beams through the implementation of matched impendence.

The concept of the gradient MS can also be introduced into acoustics to circumvent the thickness restriction imposed on conventional ACs. Recently, by using ultrathin MSs designed with transversally gradient phase[28] or impedance[29,30] profiles, novel sound manipulations on *reflected wavefronts* have been theoretically investigated. Here we focus on the acoustic MS that demonstrates *anomalous refraction* (AR) behavior for airborne sound in kilohertz regime. The flat MS is elaborately designed by arranging spatially varied subunits with coiling slits, where the elongated sound paths enable substantial phase delays. The proposed design manifests high coupling efficiency into the desired transmitted beam through a MS with a deep subwavelength thickness (~1/6.7 operational wavelength), which beats simultaneously the both limitations inherent in the conventional ACs. The redirected sound wavefronts have been successfully validated by experimental field patterns. To the best of our knowledge, so far this is the first design and experimental demonstration of the GSL-based AR phenomenon in acoustics. The present design strategy can be flexibly extended to modulate transmitted wavefronts to realize a wide variety of functionalities unattainable with conventional ACs.

# Results

**Design of the transmitted MS for airborne sound.** To efficiently steer the transmitted beam, it is necessary to introduce highly controllable and position-dependent phase shifts over the whole $2\pi$ range. In optic systems, the desired phase coverage can be readily obtained by anisotropic resonators through the cross coupling between polarizations. Unfortunately, this scheme cannot be extended to the system for airborne sound which is essentially a *scalar* wave. By resorting to building blocks made of coiling slits, recently Ref. 28 has realized the required phase profile in an ultrathin sample (and consequently demonstrated a high-quality manipulation of anomalous *reflection* wavefronts in full-wave simulations). A similar route is employed here. In a microscopic view, the coiling structure forces the sound to travel in a zigzag path and thus effectively elongates the propagation distance of sound. Note that the transmitted phase delay cannot be simply expected by the total slit length since it is determined consistently by the inherent interference among the waves traveling back and forth (due to the unavoidable impedance mismatch at the slit exits). However, the elongated path indeed gives a possibility to achieve wide phase coverage over a deep subwavelength thickness, as to be shown in Fig. 1b. In fact, zigzag channels have been extensively employed to modulate sound in applied acoustics[31]. Such folded structures are presently attracting new interest in designing metamaterials with fascinating properties[32-37], e.g. negative refractions and zero indices.

Our design strategy is described as follows. As depicted in Fig. 1a, each basic building block is integrated by a couple of vertical bars and several horizontal bars, where the air space forms zigzag slits. Specifically, the horizontal bar always starts from the top-left to form an outlet in the top-right of the subunits. This treatment provides nearly equal-distance among the neighboring outlets when different subunits are assembled together, which facilitates the practical sample design. In principle, there are many structure parameters can be tailored to attain required amplitude and phase responses. However, a comprehensive analysis on all parameters is cumbersome and beyond the scope here. In Fig. 1a all dimensions are fixed except the length *l* of the horizontal bars. As exhibited later, this variable plus the total number *n* of the horizontal bars can already provide a wide range of local amplitude and phase responses. The accuracy of design is safely guaranteed by full-wave simulations based on the finite-element method, where the solid bar is modeled as acoustically rigid with respect to air (see methods).

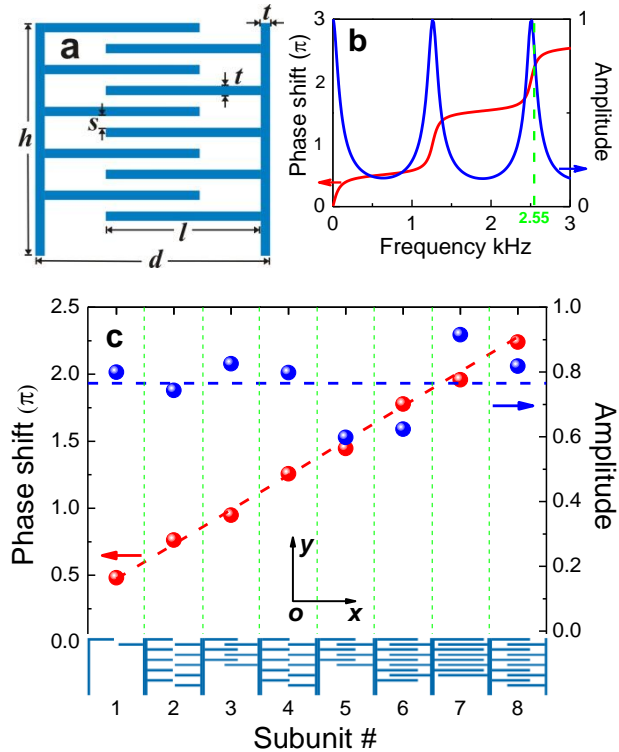

**Figure 1 | Coiling subunits and their acoustic responses.** **(a)** Schematic of the proposed coiling subunit, where the length *l* and the number *n* of the horizontal bars are adjustable parameters. **(b)** A typical set of the phase shift (red) and normalized amplitude (blue) spectra. **(c)** Phase (red) and amplitude (blue) responses varying with the numbered eight subunits optimized at 2.55kHz. Here both (a) and (b) are exemplified by the subunit #8, and the data in (b) and (c) are collected from the outgoing far-field responses for a periodical array of identical subunits, excited by a plane wave along +*y* direction.

We first calculated the sound field distribution for a periodical array of identical subunits (specified by *l* and *n*), excited by a plane wave normally onto to the structure. From the transmitted far-field the amplitude and phase shift can be extracted. As an example, in Fig. 1b we present a set of phase and amplitude spectra for a typical configuration, where the amplitude is normalized by that of incident wave. It is observed that the phase accumulation grows rapidly near the resonances and indeed covers a wide range of value over a thickness of 2cm. Here we focus on a specific frequency of 2.55kHz (corresponding to air wavelength $\lambda \approx 13.3$cm), which is selected after a full consideration of the multi-scale nature of the practical sample (see methods). For this prefixed frequency, the repeated process of numerous different configurations gives eight optimized basic building blocks, as labeled in Fig. 1c with geometry details listed in methods. In Fig. 1c we present the corresponding transmitted phase shift (red) and amplitude (blue) responses. It demonstrates that the eight discrete phase shifts cover the entire phase range and increase with a step ~$\pi/4$ among the nearest neighbors. The corresponding transmitted amplitudes are considerably large (fluctuating around 0.77, achieved by intentionally choosing configurations near resonances), which is of great benefit to high transmission efficiency. Similar to the optic cases, our MS is constructed by a one-dimensional periodical array of supercells, each formed by assembling the eight different subunits together. As shown below, thanks to the nearly constant phase gradient and amplitude profiles, such a thin MS (with thickness ~$\lambda/6.67$) controls effectively the transmitted wavefronts according to GSL, $\frac{2\pi}{\lambda}\sin\theta_t = \frac{2\pi}{\lambda}\sin\theta_i + k_a$, where $\theta_i$ and $\theta_t$ are the incident and transmitted angles, respectively, and $k_a$ is the additional momentum determined by the transversal phase gradient. The GSL equation also implies that the desired transmitted beam would become evanescent provided that the incident angle is tuned beyond a critical value $\theta_c = \arcsin(1 - 2\pi k_a/\lambda) \approx 9.6^\circ$.

Note that the current design is not a simple duplication of the coiling MS employed in Ref. 28 which aims to demonstrate anomalous reflections of sound wavefronts. Apart from removing the rigid substrate (used to produce total reflection), a crucial modification here is the elimination of the air spacing among the coiling subunits. As shown later, this treatment would significantly improve the conversion efficiency of the transmitted energy to the AR beam. Otherwise, the direct propagation of sound through the interspace will lead to a considerable contribution to the ordinary beam; this unwanted component can even dominate the transmission since the sound energy tends to transport through the straight channels directly, rather than to squeeze through narrow and long coiling slits. Another striking difference is the relaxation of the number of horizontal bars in each subunit (which is discrete and finite). This facilitates acquiring simultaneously the desired local phase and amplitude responses without incurring heavy simulation tasks.

**Numerical demonstrations.** To verify the AR behavior predicted by GSL, we first simulated a system of finite size for the prefixed operational frequency, 2.55kHz. This will provide a useful guideline in practical experiments, where the whole system could be heavily restricted by the multi-scale nature and thus the finite size effect should be understood in advance. Specifically, here a MS with length 224cm is considered, impinged normally by a Gaussian beam of width ~80cm (i.e. ~$6\lambda$).

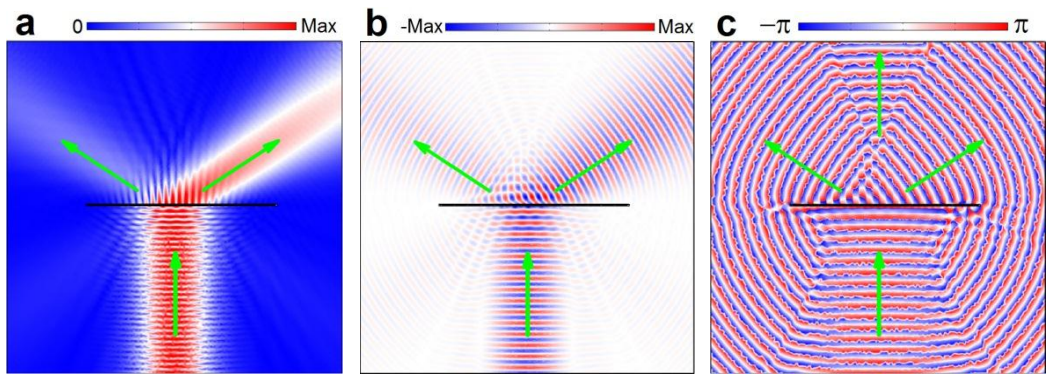

**Figure 2 | The AR behavior for a finite system.** Full-wave simulation of the sound field for a Gaussian beam of 2.55kHz impinging normally upon a MS of finite length, where **(a)**, **(b)** and **(c)** correspond to the amplitude, temporal and phase fields, respectively, with the green arrows indicating the propagation directions of wavefronts.

Figures 2a and 2b present the amplitude and temporal fields, which manifest a couple of transmitted wavefronts strikingly deflected from the incidence. The bright one propagating toward the right-hand side is exactly the desired AR beam, as predicted from the GSL with deflection angle $\theta_t \approx 56.4º$ (see arrows). This transmitted beam can be simply regarded as a consequence of constructing interference among the deep subwavelength sound sources emitted from the coiling slits. In terms of physics, the momentum mismatch between the AR beam and the incident one is compensated by the transversal gradient of the phase shifts. Due to the periodicity of the supercells arranged, the anomalous beam can also be regarded as the +1 order diffraction[38], whereas the faint beam outgoing toward the left-hand side corresponds to the −1 order one. It is of interest that the 0 order branch, i.e., the so-called ordinary refraction propagating along the incidence, is strongly suppressed. This beam, despite very weak, can be noticed in the phase pattern displayed in Fig. 2c, if away from the interfering region created by the two relatively stronger nonzero order beams. Different from the dominant +1 order branch, the weak beams of −1 order and 0 order stem mostly from the imperfect design in the phase and amplitude responses. Overall, Fig. 2 states that a field region of only several wavelengths is enough to demonstrate the AR phenomenon.

**Experimental validations.** Below we present experimental validations for the above numerical results. The sample has been fabricated by using a commercial 3D printer, which is made of plastic and behave acoustically rigid with respect to air. Fig. 3a shows a photograph of the supercell assembled by eight different subunits, which has a length 16cm, a thickness 2.0cm and a height 1.2cm. The whole sample is formed by periodically arranging a total of 7-super-cells together (more supercells used for the oblique incidence later). In experiment, the sample is tightly sandwiched between a laboratory table and a covering Plexiglass plate. For the frequency range under consideration, the parallel gap in between behaves as a waveguide and supports only a 2D propagation of sound. Absorbers are placed at the open ends of the waveguide to reduce the unwanted reflection from the free space. A Gaussian beam (of width

~60cm, i.e. ~$4.5\lambda$) is produced by a narrow microphone together with a parabolic concave-mirror[39]. The sound field behind the sample is measured by two identical microphones (of diameter ~0.7cm, B&K Type 4187): one is fixed to act as phase reference, and the other is movable to scan the field distribution behind the sample point by point. Finally, the acoustic signals are analyzed by a multi-analyzer system (B&K Type 3560B), from which both of the wave amplitude and phase can be extracted.

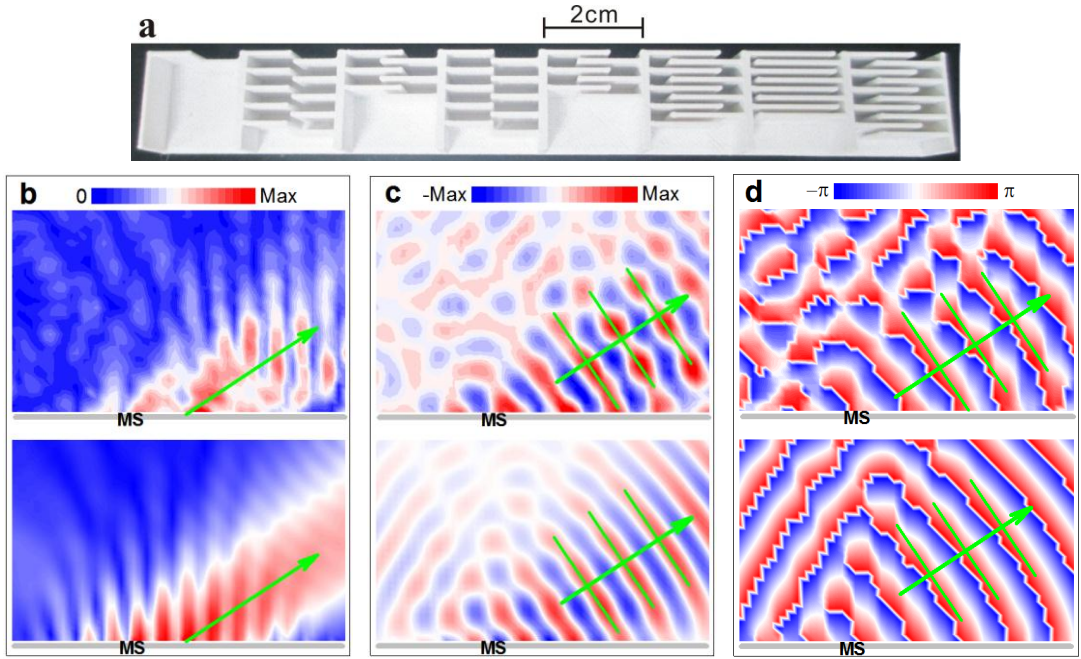

**Figure 3 | Experimental validation of the AR effect.** (**a**) Top view of the fabricated supercell. (**b**)-(**d**) Measured (upper panels) and simulated (lower panels) amplitude, temperal, and phase field distributions, respectively, excited by a normally incident Gaussian beam at 2.55kHz. The green arrows and parallel lines indicate the propagating wavefronts predicted by GSL.

In the upper panels of Figs. 3b-3d, we present the experimental amplitude, temperal, and phase fields excited by the Gaussian beam under normal incidence. The field regions displayed are $102cm \times 62cm$ (~$8\lambda \times 5\lambda$), slightly above the interface of the sample (gray). From the measured temporal and phase patterns, high-quality planar wavefronts can be observed in the bottom-right field region, associated with the notable outgoing beam in the amplitude distribution. These sound field profiles demonstrate clearly that the MS bends the sound propagation toward right-hand side, where the direction of wavefront precisely coincides with the theoretical prediction from GSL (indicated by the green arrow). For comparison, in the lower panels of Figs. 3b-3d we present the corresponding full-wave simulations similar to Fig. 2, but with a shorter sample and a narrower Gaussian beam (same as in the experiment). It is observed that the measured sound field profiles agree excellently well with the full-wave simulations, especially in the region of high amplitudes that demonstrate the AR behavior. The difference observable in the weak field region may come from the unavoidable measurement noise or reflection from the boundary.

**Performance evaluations.** In Fig. 4a we present the numerical transmission spectra for an idealized system, i.e. an infinite array of supercells. It is observed that the total transmission is considerably high (~80%) around the designed frequency (2.55kHz). To further evaluate the conversion efficiency to the AR beam, the transmission is rigorously decomposed into its diffractive components by implementing Fourier transform of the transmitted field. Within this frequency range, only three diffractive beams are allowed, i.e. $-1$, 0 and $+1$ orders. As shown in Fig. 3a, most of the transmitted energy is converted to the beam of $+1$ order, i.e. the desired AR beam predicted by GSL, associated with much less energy coupled into the other two. In particular, the energy transported through the so-called ordinary refraction (i.e. 0 order) turns even negligible near 2.55kHz, as consistent with Fig. 2. This is because that the wave energy can only be funneled through the slits, which will produce the single AR beam as predicted from GSL. Therefore, the transmitted component of the ordinary refraction stems only from the imperfect design of the MS, e.g. the fluctuating amplitude responses and the inevitable near-field coupling among the subunits. This is different from many designs in optics where the wave energy can directly penetrate through the dielectrics supporting metallic resonators, leading to a considerable contribution to the ordinary beam.

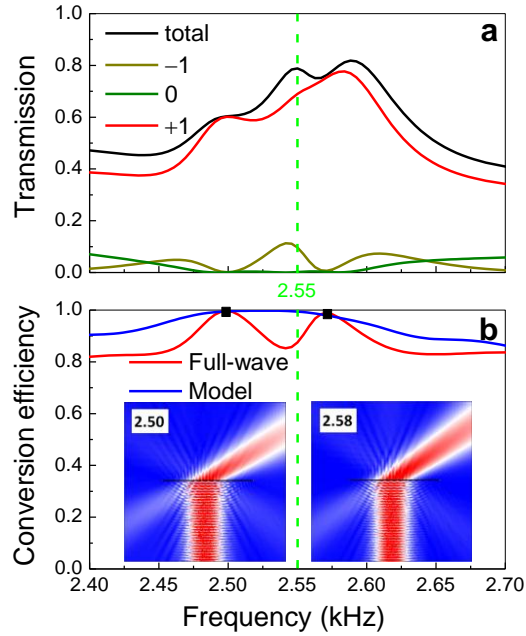

**Figure 4 | Frequency broadening and high-efficiency conversion.** (**a**) Power transmission spectrum for a plane wave incident normally to the MS made of infinite supercells, accompanying with its three diffractive components. (**b**) Energy conversion efficiency of the desired AR beam, obtained from either the full-wave simulation or the simplified model. Insets: simulated amplitude fields excited by a Gaussian beam at the frequencies of maximum efficiencies (denoted by squares).

In Fig. 4b the red line shows the frequency dependent conversion efficiency for the AR beam, defined by the energy ratio to the total transmission. It is considerably high (>80%) over the whole frequency range under consideration, although the design is optimized for a specific frequency. The frequency broadening effect has also been verified well in experiments. To roughly explain this behavior, we have studied the frequency dependences of the phase and amplitude responses for the eight subunits individually (similar to Fig. 1b). Within this frequency range, overall, the phase shifts cover a full $2\pi$ span and exhibit a trend of monotonous increase from the subunit #1 to #8, consistently leading to positive (despite nonuniform) transversal momentums. So it is the average effect (over all subunits) that results in the broadening of operating frequency. This qualitative picture is further tested by a simple model based on the Huygens-Fresnel principle: the slit-exits are approximated as subwavelength-sized point sources and assigned orderly with the simulated phase shifts and amplitudes. The radiation ratio of the desired AR beam can be extracted straightforwardly from the sound field superposed by such an array of point sources. As manifested by the blue line, high performance indeed covers a wide frequency range, associated with almost perfect conversion near the designed 2.55kHz. Comparatively, in the full-wave simulation, the maximum conversion efficiencies (exceeding 98%) slightly deviate from the prescribed frequency and occurs at 2.50kHz and 2.58kHz. As displayed in the insets, for both frequencies the unwanted −1 order transmitted beam is considerably reduced with respect to 2.55kHz (see Fig. 2a). This improvement can be attributed to the unavoidable coupling effect among the different subunits.

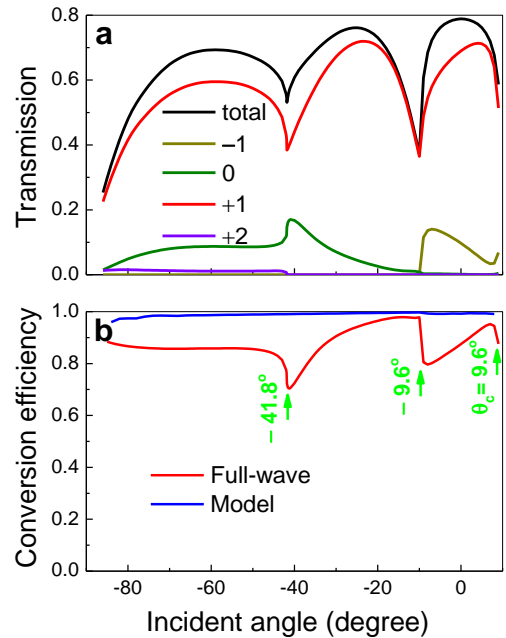

**Figure 5 | Broad-angle behavior.** (**a**) The angular dependence of the power transmission for 2.55kHz, accompanying with its four diffractive components. (**b**) Energy conversion efficiency of the desired AR beam, obtained from either the full-wave simulation or the simplified model. Here the green arrows indicate the angles of Wood anomalies, and $\theta_c$ represents the critical angle where the AR (i.e. +1 order) beam vanishes.

The angular robustness of the AR effect is highly desirable for the further realization of relevant devices (e.g. focusing lenses), based on the primary design starting from the normal incidence. In Fig. 5a we present the power transmission over a wide range of incident angles, together with its diffractive components. (Note that the +2 order diffractive beam, despite small, appears as the incident angle $\theta_i < -41.8^\circ$.) It is observed that as a whole the total transmission is considerably high, where the major contribution comes from the desired AR beam. This leads to the angularly robust high conversion efficiency in Fig. 5b (red line). Physically, this broad-angle effect stems from the extreme anisotropy of the subunits: sound can only propagate through the slits and there is no direct transversal coupling among different slits (except via the exits). Therefore, both the transmitted amplitude and phase responses vary softly with incident angles. Again, the high conversion can be roughly understood from the simple model (see blue line). The difference turns remarkable near the angles associated with Wood's anomalies (see green arrows), because of the increasing coupling among the different subunits in the real acoustic MS.

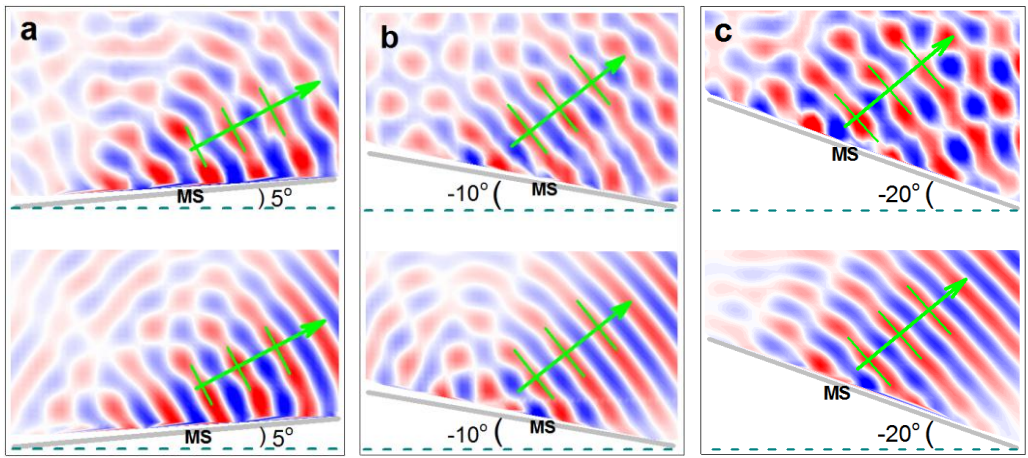

**Figure 6 | Experimental validations for oblique incidences.** Experimental (upper panels) and numerical (lower panels) temperal field distributions excited by a Gaussian beam of 2.55kHz along +$y$ direction, where (**a**), (**b**) and (**c**) correspond to the samples (gray) with tilted angles of $5^\circ$, $-10^\circ$ and $-20^\circ$, respectively. The green arrows and parallel lines indicate the propagating wavefronts predicted by GSL.

To realize oblique incidence in the experiment, it is convenient to tilt the sample with respect to the whole fixed measurement system. Here we present examples for the incident angles $\theta_i = 5°$, $-10°$ and $-20°$, where the first one closes to the critical angle $\theta_c$, and the last one corresponds to a relatively large angle allowed in the measurement. The instant pressure fields are deomonstrated in Fig. 6, where the upper and lower panels correspond to the experimental data and numerical comparisons, respectively. Again, for each case the sound field pattern displays close resemblance between the measured and simulated results: the dominant diffraction comes from the AR beam predicted by GSL (as indicated by the green arrow).

**Discussion**

Note that in our design the sample thickness is determined by the subunit supporting the largest phase delay. Qualitatively, a periodical array of the coiling subunits can be viewed as a thin layer of effective medium with high refractive index, such that the transmitted sound could be heavily delayed through it. In spite of this, the effective medium is not replaceable by the natural solid with low sound speed, e.g. the soft rubber. This is explained as follows. For the effective medium, roughly, the bulk modulus $\kappa_e$ can be estimated from its filling ratio, and the refractive index $n_e$ can be estimated by the elongated slit length over the layer thickness. For example, the effective parameters (scaled by those of air) for the eighth configuration employed here $\kappa_e \sim 1.5$ and $n_e \sim 7$, which further give rise to the effective mass density $\rho_e \sim 75$ and impedance $Z_e \sim 10$. As shown in Fig. 1b, the moderate $Z_e$ provides relatively wide resonances over a considerable transmission background, which benefits to the performance of MS. In fact, the impedance matching could be further improved by optimizing more geometric parameters, such as introducing a spatial gradient of the length for horizontal bars in each subunit. In contrast, the impedance of the natural solid is usually two to three orders higher than air due to the great density ratio. Besides the strong impedance contrast, another notable reason (to exclude natural solids) is the huge transversal impedance that arises from the impenetrability through the vertical bars. The extreme anisotropy enables almost invariant phase delay for different incident angles, since the wave energy can only be transported along the thickness direction of the sample (associated with angularly invariant propagating distance). This guarantees the effectiveness of the broad-angle AR designed only from the normal incidence, and also ensures further realization of arbitrary wavefront manipulations based on a similar design. Also, it is worth noting that the MS proposed differs from the well-explored airborne gratings[40-42], in which the extraordinarily high transmission is accomplished by a thickness comparable with wavelength, accompanying with much less capability in manipulating sound wavefronts.

In conclusion, we have demonstrated an effective design on the AR effect of airborne sound through an ultrathin MS (~1/6.7 operational wavelength). By elaborately optimizing the subunit geometries, the proposed flat MS exhibits excellent performance: high conversion efficiency over a broad range of frequencies and incident angles. The measured sound field patterns exhibit high-quality redirected wavefronts and agree well with those predicted from full-wave simulations. In principle, a similar design can be extended to the 3D case (by using coiled hole-arrays) to achieve arbitrary 3D shaping of wavefronts, such as in generating acoustic vortices with well-defined orbital angular momentums and non-diffracting Bessel beams. This study may pave the way to significant advances in steering transmitted wavefronts by compact acoustic elements.

**Methods**

**Simulations.** Throughout the paper, all full-wave simulations are accurately performed based on the commercial finite-element solver (COMSOL Multiphysics), in which the sound speed 340m/s is employed for the practical room temperature of ~15°C. In the simulations, all microstructures with actual geometric sizes are fully considered. The plastic frame of the coiling MS is modeled as acoustically rigid. (According to the well-known mass-density law, the transmission through a plastic plate of thickness 0.8mm, i.e., the smallest thickness involved here, can be estimated as low as 0.004 around 2.55kHz.) Besides, we ignore the dissipation that mainly comes from the viscosity of air within a thin layer near the channel surface. It will not considerably damage the predicted phenomena for the channel size and wavelength involved here. In fact, similar system parameters have been widely employed in the coiling metamaterials and the other holey structures[11,13,34,35]. Except for the periodical boundary condition applied in the specified cases, the radiation boundary condition is set for the remaining situations. The total power transmission is calculated by integrating the Poynting vectors and normalized to the incidence. The relative weights for varied diffractive branches are extracted after precisely calculating the scattering matrix of the complex sample.

**Sample preparation.** In the procedure of sample design, a practical issue originated from the multi-scale nature of the whole experimental system must be fully taken into account. There are four length scales involved in descending order: the total length of sample, the wavelength, the size of subunit, and the size of microstructure in each subunit. In our experiment, the maximum feature size, i.e. the sample length, is limited by the size of laboratory table (150cm x 300cm), and the minimum feature size, i.e. the thickness of the horizontal bars, is determined by the manufacture accuracy (~0.1mm). A comprehensive assessment of them leads to the currently used wavelength (~13.3cm), and the sample geometry, i.e. the thickness ($h$) and length ($d$) of the subunit $h = d = 2.0 \text{cm}$, the thickness ($t$) and spacing ($s$) of the horizontal bars $t = 0.8 \text{mm}$ and $s = 1.0 \text{mm}$, respectively. Besides the specified geometries, each subunit is featured by two tunable parameters: the number ($n$) and the length ($l$) of the horizontal bars. The former is manifested directly in the inset of Fig. 1b. The latter for the eight subunits are orderly listed as follows: 8.4, 8.9, 11.4, 9.5, 12.4, 14.2, 17.1, and 13.2 millimeters. The sample (glued with many supercells) is fabricated with thermo-plastics via 3D printing technique, where a supercell is finished in a single printing.

**References:**

1. Cervera, F. *et al.* Refractive acoustic devices for airborne sound. *Phys. Rev. Lett.* **88**, 023902 (2002).

2. Qiu, C., Zhang, X. & Liu, Z. Far-field imaging of acoustic waves by two-dimensional sonic crystal. *Phys. Rev. B* **71**, 054302 (2005).

3. Liang, F. *et al.* Acoustic backward-wave negative refractions in the second band of a sonic crystal. *Phys. Rev. Lett.* **96**, 014301 (2006).

4. Lu, M. H. *et al*. Negative birefraction of acoustic waves in a sonic crystal. *Nature Mater.* **6**, 744-748 (2007).

5. Wang, Y., Deng, K., Xu, S., Qiu, C., Yang, H. & Liu, Z. Applications of antireflection coatings in sonic crystal-based acoustic devices. *Phys. Lett. A* **375**, 1348 (2011).

6. Yang, Z., Mei, J., Yang, M., Chan, N. H. & Sheng, P. Membrane-type Acoustic metamaterial with negative dynamic mass. *Phys. Rev. Lett.* **101**, 204301 (2008).

7. Lee, S. H., Park, C. M., Seo, Y. M., Wang, Z. G. & Kim, C. K. Composite acoustic medium with simultaneously negative density and modulus. *Phys. Rev. Lett.* **104**, 054301 (2010).

8. Yang, M., Ma, G., Yang, Z. & Sheng, P. Coupled membranes with doubly negative mass density and bulk modulus. *Phys. Rev. Lett.* **110**, 134301 (2013).

9. Park, J. J., Lee, K. J. B., Wright, O. B., Jung, M. K. & Lee, S. H. Giant acoustic concentration by extraordinary transmission in zero-mass metamaterials, *Phys. Rev. Lett.* **110**, 244302 (2013).

10. Fleury, R. & Alu, A. Extraordinary sound transmission through density-near-zero ultranarrow channels. *Phys. Rev. Lett.* **111**, 055501 (2013).

11. Li, J., Fok, L., Yin, X., Bartal, G. & Zhang, X. Experimental demonstration of an acoustic magnifying hyperlens. *Nature Mater.* **8**, 931–934 (2009).

12. Liu, F., Cai, F., Peng, S., Hao, R., Ke, M. & Liu, Z. Parallel acoustic near-field microscope: A steel slab with a periodic array of slits. Phys. Rev. E **80**, 026603 (2009).

13. Zhu, J. *et al*. A holey-structured metamaterial for acoustic deep-subwavelength

imaging. *Nature Phys.* **7**, 52–55 (2011).

14. Zhang, S., Yin, L. & Fang, N. Focusing ultrasound with an acoustic metamaterial network. Phys. Rev. Lett. **102**, 194301 (2009).

15. Christensen, J. & Garcia de Abajo, F. J. Anisotropic Metamaterials for Full Control of Acoustic Waves. *Phys. Rev. Lett.* **108**, 124301 (2012).

16. García-Chocano, V. M., Christensen, J. & Sánchez-Dehesa, J. Negative refraction and energy funneling by hyperbolic materials: an experimental demonstration in acoustics. *Phys. Rev. Lett.* **112**, 144301 (2014).

17. D'Aguanno, G. *et al*. Broadband metamaterial for nonresonant matching of acoustic waves. *Sci. Rep.* **2**, 340 (2012).

18. Popa, B.-I., Zigoneanu, L. & Cummer, S. A. Experimental acoustic ground cloak in air. *Phys. Rev. Lett.* **106**, 253901 (2011).

19. Sanchis, L. *et al*. Three-dimensional axisymmetric cloak based on the cancellation of acoustic scattering from a sphere. *Phys. Rev. Lett.* **110**, 124301 (2013).

20. Zigoneanu, L., Popa, B.-I. & Cummer, S. A. Three-dimensional broadband omnidirectional acoustic ground cloak. *Nature Mater.* **13**, 352 (2014).

21. Yu, N., Genevet, P., Kats, M. A., Aieta, F., Tetienne, J.-P., Capasso, F. & Gaburro, Z. Light propagation with phase discontinuities: generalized laws of reflection and refraction. *Science* **334**, 333–337 (2011).

22. Ni, X., Emani, N. K., Kildishev, A. V., Boltasseva, A. & Shalaev, V. M. Broadband Light Bending with Plasmonic Nanoantennas. *Science* **335**, 427 (2012).

23. Sun, S. et al. Gradient-index meta-surfaces as a bridge linking propagating waves and surface waves. *Nature Mater.* **11**, 426–431 (2012).

24. Grady, N. K. *et al*. Terahertz metamaterials for linear polarization conversion and anomalous refraction. *Science* **340**, 1304 (2013).

25. Kildishev, A. V., Boltasseva, A. & Shalaev, V. M. Planar photonics with metasurfaces. *Science* **339**, 1232009–1232009 (2013).

26. Pfeiffer, C. & Grbic, A. Metamaterial Huygens' surfaces: tailoring wave fronts with reflectionless sheets. *Phys. Rev. Lett.* **110**, 197401 (2013).

27. Monticone, F., Estakhri, N. M. & Alù, A. Full control of nanoscale optical

transmission with a composite metascreen. *Phys. Rev. Lett.* **110**, 203903 (2013).

28. Li, Y., Liang, B., Gu, Z., Zou, X. & Cheng. J. Reflected wavefront manipulation based on ultrathin planar acoustic metasurfaces. *Sci. Rep.* **3**, 2546 (2013).

29. Zhao, J., Li, B., Chen, Z. & Qiu, C. W. Manipulating acoustic wavefront by inhomogeneous impedance and steerable extraordinary reflection, *Sci. Rep.* **3**, 2537 (2013).

30. Zhao, J., Li, B., Chen, Z. & Qiu, C. W. Redirection of sound waves using acoustic metasurface. *Appl. Phys. Lett.* **103**, 151604 (2013).

31. Rossing, T. D., Moore, R. F. & Wheeler, P. A. *The Science of Sound* (Addison Wesley, San Francisco, 2002).

32. Liang, Z. & Li, J. Extreme acoustic metamaterial by coiling up space. *Phys. Rev. Lett.* **108**, 114301 (2012).

33. Li, Y. *et al*. Acoustic focusing by coiling up space. *Appl. Phys. Lett.* **101**, 233508 (2012).

34. Liang, Z. *et al*. Space-coiling metamaterials with double negativity and conical dispersion. *Sci. Rep.* **3**, 1614 (2013).

35. Xie, Y., Popa, B.-I., Zigoneanu, L. & Cummer, S. A. Measurement of a broadband negative index with space-coiling acoustic metamaterials. *Phys. Rev. Lett.* **110**, 175501 (2013).

36. Xie, Y., Konneker, A., Popa, B.-I. & Cummer, S. A. Tapered labyrinthine acoustic metamaterials for broadband impedance matching. *Appl. Phys. Lett.***103**, 201906 (2013).

37. Frenzel, T., Brehm, J. D., Bückmann, T., Schittny, R., Kadic, M. & Wegener, M., Three-dimensional labyrinthine acoustic metamaterials. *Appl. Phys. Lett.* **103**, 061907 (2013).

38. Larouche, S. & Smith, D. R. Reconciliation of generalized refraction with diffraction theory. *Opt. Letts.* 37, 2391-2393 (2012).

39. Lu, J., Qiu, C., Xu, S., Ye, Y., Ke, M. & Liu, Z. Dirac cones in two-dimensional artificial crystals. *Phys. Rev. B* **89**, 134302 (2014).

40. Christensen, J., Fernandez-Dominguez, A. I., De Leon-Perez, F., Martin-Moreno,

L. & Garcia-Vidal, F. J. Collimation of sound assisted by acoustic surface waves. *Nature Phys.* **3**, 851–852 (2007).

41. Lu, M. H. *et al*. Extraordinary acoustic transmission through a 1D grating with very narrow apertures. *Phys. Rev. Lett*. **99**, 174301 (2007).

42. Christensen, J., Martin-Moreno, L. & Garcia-Vidal, F. J. Theory of resonant acoustic transmission through subwavelength apertures. *Phys. Rev. Lett.* **101**, 014301 (2008).

**Acknowledgements**

This work is supported by the National Natural Science Foundation of China (Grant Nos. 11174225, 11004155, 11374233, and J1210061); Open Foundation from State Key Laboratory of Applied Optics of China, and the Program for New Century Excellent Talents (NCET-11-0398).

**Author contributions**

Z.L. and C.Q. conceived the original idea. K.T. performed the simulations and measurements with the help of J.L. and Y.Y. C.Q. and Z.L. supervised the theory and simulations; C.Q. and M.K. supervised the experimental measurements. C.Q. wrote the manuscript together with Z.L. and K.T. All authors contributed to scientific discussions of the manuscript.

**Competing financial interests**

The authors declare no competing financial interests.